\documentclass[a4paper,11pt]{article}
\pdfoutput=1 % if your are submitting a pdflatex (i.e. if you have
\usepackage{jcappub} % for details on the use of the package, please
\usepackage{wasysym}
\usepackage[T1]{fontenc}
\usepackage{lmodern} 

\title{\boldmath Exploring 0.1-10\,eV axions with a new helioscope concept. }

\author{J.~Gal\'an$^a$, T. Dafni$^a$, E. Ferrer-Ribas$^b$, I. Giomataris$^b$, F.J. Iguaz$^a$, I.G. Irastorza$^a$, J.A. Garc\'ia$^a$, J. Gracia$^a$, G. Luzon$^a$, T.~Papaevangelou$^b$, J.~Redondo$^a$, A.~Tom\'as$^a$\,\footnote{Present address : High Energy Physics Group, The Blackett Laboratory, Imperial College, London, UK} }

\affiliation[a]{Laboratorio de F\'isica Nuclear y Astropart\'iculas, Universidad de Zaragoza, Zaragoza, Spain}
\affiliation[b]{IRFU, Centre d'\'Etudes Nucl\'eaires de Saclay (CEA-Saclay), Gif- sur-Yvette, France}

\emailAdd{javier.galan.lacarra@cern.ch}

\abstract{We explore the possibility to develop a new axion helioscope type, sensitive to the higher axion mass region favored by axion models. We propose to use a low background large volume TPC immersed in an intense magnetic field. Contrary to traditional tracking helioscopes, this detection technique takes advantage of the capability to directly detect the photons converted on the buffer gas which defines the axion mass sensitivity region, and does not require pointing the magnet to the Sun. The operation flexibility of a TPC to be used with different gas mixtures (He, Ne, Xe, etc) and pressures (from 10\,mbar to 10\,bar) will allow to enhance sensitivity for axion masses from few meV to several eV. We present different helioscope data taking scenarios, considering detection efficiency and axion absorption probability, and show the sensitivities reachable with this technique to be few $\times$ 10$^{-11}$ GeV$^{-1}$ for a 5\,T, m$^3$ scale TPC. We show that a few years program taking data with such setup would allow to probe the KSVZ axion model for axion masses above $\gtrsim$\,100\,meV.} 

\begin{document}
\maketitle
\flushbottom

\section{Introduction}
\label{sec:intro}

The axion is a hypothetical neutral pseudoscalar particle which was already predicted in 1977~\cite{weinberg, Wilczek, PQ0, Kim}. This weakly interacting particle came out as a simple solution to the CP problem of strong interactions in QCD~\cite{PQ1,PQ2}. Two main theoretical approaches are considered today as a model to define the properties of the axion. The KSVZ model assumes only pure hadronic interactions~\cite{KSVZ1,KSVZ2}, while the DFSZ model allows axion-fermion interactions at tree level~\cite{DFSZ1,DFSZ2}. These two models constrain the relation between the axion mass and its coupling to ordinary matter. If the mass were too high the coupling to ordinary matter would be so intense that it should have been already observed by previous experiments, this massive axion was easily excluded~\cite{PQ1,weinberg,Wilczek}. Thus the axion mass and coupling must be low, and it is commonly known as invisible axion. The particular properties of the axion can be restricted by the actual observational consequences that its existence would imply in astrophysics and cosmology~\cite{Raffelt0, Sikivie0}. Recently, strong bounds on the axion-photon coupling, $g_{a\gamma}$, derived from globular clusters sets an upper limit for the axion interaction of $g_{a\gamma} \textless 0.66 \times 10^{-10}$\,GeV$^{-1}$~\cite{Mirizzi}. Moreover, the fact that the axion would have a non-vanishing mass places it as a good dark matter candidate under some conditions~\cite{Dine,Abbot,Turner,Raffelt1}. The idea of the existence of the axion is particularly attractive because of the way it brings cosmology, astrophysics and particle physics together so elegantly. 

\medskip
The Primakoff effect describes the conversion of photons into axions in the presence of virtual photons provided by external electric or magnetic fields. These fields can be the Coulomb field of a nucleus, the electric field of charged particles in the hot plasma of stars or a magnetic field in the laboratory. The inverse process, in which an axion is converted into a photon, provides the basic principle for most axion search experiments. First experiments to search for the invisible axion were suggested by Sikivie~\cite{Sikivie1} in 1983, by using an external magnetic field to convert axions to photons. The existence of the axion would not only entail a theoretical breakthrough, but from a experimental point of view, it would also provide us with new tools to study the Sun and the galaxy, and observe our universe with new optics.

We can distinguish between three type of experiments that search for axions; those from galactic origin (axion haloscopes), those searching axion production in the Sun (axion helioscopes), and also pure laboratory experiments, in which the axions are produced by lasers in strong magnetic fields (for a recent review on axion searches see~\cite{questAxions}). The present work belongs to the helioscope search experiments. 

\section{Helioscope searches}

\begin{figure}[htbp]
\begin{center}
\includegraphics[width=10cm]{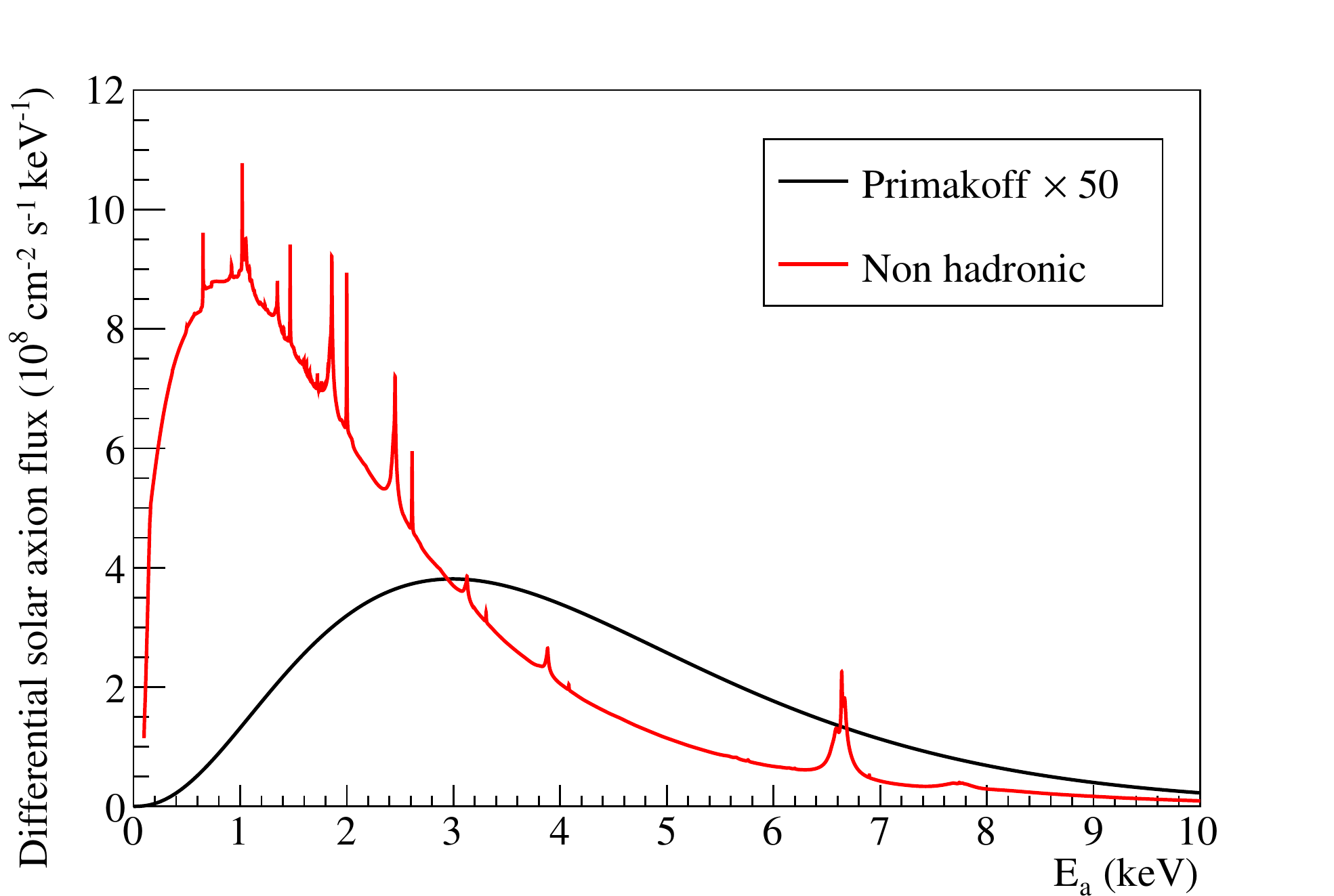}
\caption{Axion flux at Earth by pure-hadronic (KSVZ) axions produced by the Sun core for an axion-photon coupling of $g_{a\gamma}$ = 10$^{-12}$GeV$^{-1}$ (flux scaled by a factor $\times$50 for comparison). And flux produced by DFSZ axions model where axions have tree level interactions with fermions for an axion-electron coupling of $g_{ae}$ = 10$^{-13}$GeV$^{-1}$.}
\label{fig:axionFlux}
\end{center}
\end{figure}

If axions exist they must be produced in the core of the Sun through Primakoff effect for purely hadronic axions (KSVZ), and additionally through axion-electron processes if axions couple to electrons at tree level (DFSZ). The very well-known temperature and density of the Sun core given by the Standard Solar Model allows us to calculate accurately the axion flux expected on Earth (See~\cite{axionFlux} and references therein). The detection of these axions~(see Figure~\ref{fig:axionFlux}) is the basis for any helioscope experiment.

There are basically two main solar axion helioscope search lines. The first line was exploiting the conversion of axions into photons by the electromagnetic field of a regular (crystalline) atomic lattice. The main idea is to use a regular lattice crystal to enhance axion-photon conversion for the different Bragg diffraction energies given by the different atomic plane distances. The angle of incidence of these axions entering the crystal would produce a very particular daily pattern that would allow to validate an axion signal~\cite{HoogeveenBragg,PaschosBragg, CreswickBragg}. Several low background crystals installed at underground laboratories have taken data, providing limits to the axion-photon coupling in the order of $\lesssim2\cdot10^{-9}$GeV$^{-1}$~\cite{SOLAX, COSME, DAMA, CDMS}. However, the prospects for this detection technique present some limitations~\cite{prospectsCristals,prospectsAvignone}, and thus it is not competitive with the currently running helioscope experiments.

\medskip
Today, the most extended technique for helioscope searches is the \emph{tracking}\footnote{We introduce the term \emph{tracking} in this manuscript to differentiate from the helioscope concept we will present in this work. Although in the existing literature this term is not frequently used.} helioscope idea~\cite{Sikivie1}. An axion propagating inside a magnetic field interacts only with the transversal component of the field. This experiment consists of an intense magnetic field which is capable to track the Sun in such a way that the field component is kept transversal to the propagation of axions coming from the Sun. A first experiment using this approach was already carried out at Brookhaven in 1992, using a stationary magnet~\cite{Lazarus}. This first experiment was seeking an axion signal by using a magnet connected to an x-ray detector for a few minutes during the day, when the Sun was aligned with the transversal direction of the field. A major advance on this detection technique was carried out by using a magnet with longer tracking capabilities by making use of a moving platform~\cite{Sumico1,trackingHelioscope}. The idea we will present is closely related to the tracking helioscopes detection technique, therefore we provide few details that will support the description of our proposal in section~\ref{sec:modHeli}. The tracking helioscope is so far the most sensitive helioscope experiment, thus we will also present a state of the art and prospects using this technique.

%\subsection{Tracking helioscope detection principle and searches}

\subsection{The tracking helioscope detection technique}
\label{sec:tckHelioscope}

The tracking helioscope consists of a long vacuum pipe (magnet bore) where the intense magnetic field is confined. The axion converts to a photon of the same momentum as the incoming axion by interacting with the virtual photon provided by the magnetic field. Since the momentum transfer must be very low, the photon produced preserves the original energy of the axion and continues travelling without modification of the original axion trajectory. The extremes of the pipe are covered with low background x-ray detectors that detect this photon, typically in the 1-10 keV range (see Figure~\ref{fig:axionFlux}).

\medskip
The sensitivity of the experiment depends directly on the probability of conversion of an axion to photon inside the magnetic field. This quantity depends on the magnetic field intensity of the transversal component, $B$, the length of the magnet, $L$, and the coupling of axion to photon, $g_{a\gamma}$, by the following expression~\cite{raffeltAxion},

\begin{equation}\label{eq:convVacuum}
P_{a\gamma} = (g_{a\gamma}B/2)^2 L^2.
\end{equation}

\begin{figure}[h!]
\begin{center}
\includegraphics[width=13cm]{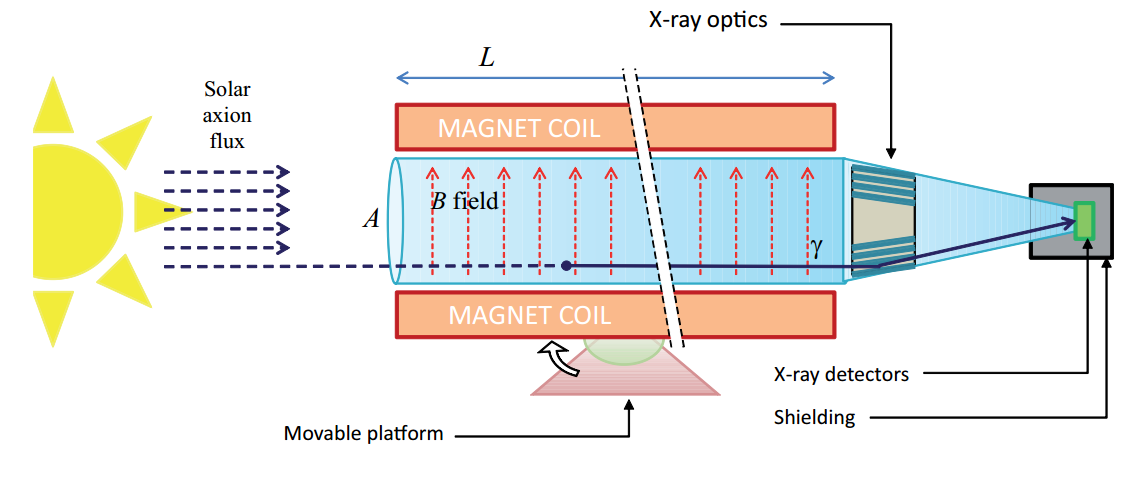}
\caption{ A conceptual drawing of a tracking helioscope magnet, including a moving platform allowing to follow the Sun, a magnet bore where axions are converted to photons, and x-ray focusing device to enhance the signal-to-noise ratio in the x-ray detectors placed at the end of the magnet.}
\label{fig:trackingHelioscope}
\end{center}
\end{figure}

The number of photons converted are integrated over the exposure time, and the cold bore aperture area. The photons detected during tracking must be well above the background level of the detectors. Longer exposure times allow to reduce uncertainties on the detectors background level and increase the possibility to observe a counting excess over the natural background. Thus, the background level of the detector becomes a relevant quantity on the final sensitivity of the experiment. In tracking helioscopes the effect of background is minimized by shielding the detector against external radiation and by focusing the signal in a reduced area of the detector by using x-ray optics (see Figure~\ref{fig:trackingHelioscope}).

\medskip
This technique only allows searching for axions masses below $m_a\lesssim\sqrt{4\pi E_a/L}$ (for a 10\,m magnet and an axion energy of $E_a=$\,4.2\,keV this implies $m_a \lesssim$ 20\,meV). For higher masses, the coherence is lost and the momentum transfer, $q$, starts to be non-negligible. A way to recover coherence and extend this kind of searches to higher masses consists in filling the cold bore with a buffer gas, thus providing the photon with an effective mass, $m_\gamma$. The conversion probability in a gaseous medium is given by~\cite{vanBibber},

\begin{equation}\label{eq:conversion}
P_{a\gamma} = \frac{(g_{a\gamma}B/2)^2}{q^2+\Gamma^2/4} \left[ 1 + e^{-\Gamma L} - 2e^{-\Gamma L/2} \mbox{cos}(qL) \right],
\end{equation}

\noindent where $\Gamma$ is the absorption coefficient of the buffer gas, and $q$ is the momentum transfer now given by the effective photon mass, $m_\gamma$, the axion energy, $E_a$ , and the axion mass, $m_a$, $q = (m_a^2 -m_\gamma^2)/2E_a$. The photon mass, $m_\gamma$, depends on the atomic scattering factor of the buffer gas, $f_1$, and its density, $\rho$, 

\begin{equation}\label{eq:mgamma}
m_\gamma^2 = 4\pi r_o (N_A/A m_u) \rho f_1
\end{equation}

\noindent where $N_A$ is the Avogadro number, $A$ is the atomic number, $m_u$ is the atomic mass unit, and $r_o$ the electron classical radius.

\medskip
The conversion probability given in equation~(\ref{eq:conversion}) is in practice the probability that a photon coming from axion-photon conversion inside the buffer gas reaches the end of the bore. In order words, it already includes the absorption of gammas in the buffer gas after axion-photon conversion\footnote{We will redefine this quantity as $P_{a\gamma} \equiv P^T_{a\gamma}$ to differentiate with the definition of conversion probability used later.}. In this type of helioscope the number of photons re-absorbed in the buffer gas are lost, and thus the use of low $\Gamma$ gases is preferred.
%it is optimum to use as low absorption gas as possible, since the number of photons re-absorbed (or lost) in the buffer gas will increase for higher values of $\Gamma$. Thus, a lower number of photons would reach the detectors at the end of the bore reducing the sensitivity of the experiment.

\subsection{Tracking helioscope searches and prospects}

The CERN Axion Solar Telescope (CAST) collaboration has used this technique to obtain the best sensitivity limit in a wide axion mass range~\cite{CAST7,CAST6,CAST5,CAST4,CAST3,CAST2,CAST1}. CAST uses a dipole magnet 9.6\,m long with an intense magnetic field of 8.9\,T, and 3 hours tracking capability per day. The experiment was divided in 2 phases; a vacuum phase and a buffer gas phase. The buffer gas phase was at the same time divided in two periods, one period was running with $^4$He, and the other period was using $^3$He, allowing to scan masses up to 1\,eV. During the second phase, CAST was only sensitive to a narrow axion mass that depends on the gas buffer density. The buffer gas density was smoothly increased in incremental pressure steps which overlapped in order to have a continous axion mass coverage, increasing considerably the data taking period required compared to the CAST vacuum phase. Previously, the Tokio helioscope reported also on axion searches performed with a shorter length and lower field intensity magnet. Despite this, the shorter magnet length allowed them to perform the first axion helioscope search for masses $\gtrsim 1$\,eV~\cite{Sumico2,Sumico3}.

\begin{figure}[ht!]
\begin{center}
\includegraphics[width=12cm]{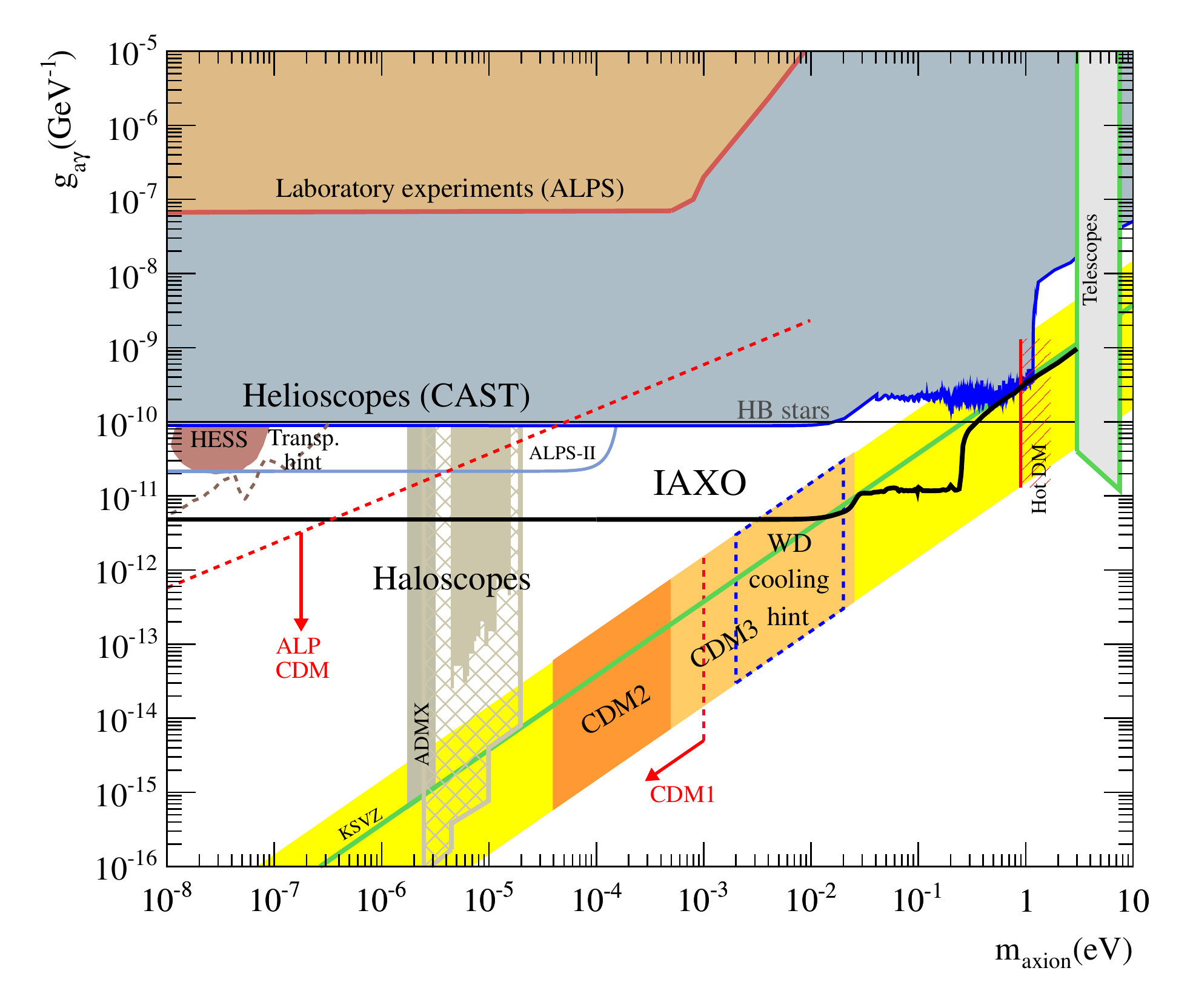}
\caption{The axion-photon coupling versus the axion mass parameter space showing the actual (CAST) and future helioscope (IAXO) sensitivities, together with resonant cavities (ADMX) searches and prospects. See~\cite{questAxions} for a detailed explanation on astrophysical and cosmological limits, and axion searches. }
\label{fig:iaxo_prospects}
\end{center}
\end{figure}

\medskip
The IAXO collaboration is developing the future generation tracking helioscope magnet~\cite{IAXO1,IAXO2}. IAXO will built a dedicated 8-bores large-aperture superconducting magnet 20\,m long, reaching an average field intensity of 2.5\,T. Each of the 8 (60\,cm diameter) magnet bores will be equipped with x-ray focussing devices that will allow to focus the large aperture area in a spot of just 0.2\,cm$^2$ allowing to increase significatively the signal-to-background ratio. A dedicated tracking system will allow to take data during 12\,h per day. All these enhancements will allow IAXO to improve by 4-5 orders of magnitude the sensitivity of CAST in terms of signal-to-noise, reaching sensitivities of a few $\times$10$^{-12}$GeV$^{-1}$. IAXO would allow us to prove axions in an extended QCD axion favored region not previously accessed by any running or proposed experiment (see Figure~\ref{fig:iaxo_prospects}).

We will show that the new helioscope type we propose in this work will allow us to cover a new theoretically favored part of the axion parameter space not explored so far. This detection technique will tend to be more efficient than tracking helioscopes for higher axion masses, for which higher buffer densities are required. In tracking helioscopes the signal is diminished for higher gas densities due to photon re-absorption. However, the helioscope type we will propose will directly detect photons absorbed in the buffer gas, making it more efficient for higher buffer gas densities, $\rho$ and gases with higher absorption coefficient, $\Gamma$. 

%The Axion Modulation hELIoscope Experiment (AMELIE) will also allow us to cover a favored theoretical region. The new axion detection technique that will be presented in the next section will tend to be more efficient than tracking helioscopes when we go towards higher axion masses.

\section{A new helioscope detection technique}
\label{sec:modHeli}

\begin{figure}[ht!]
\begin{center}
\includegraphics[width=10cm]{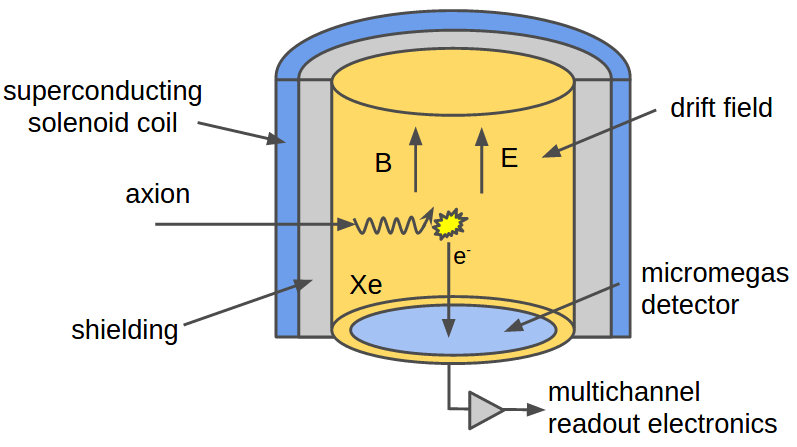}
\caption{ A conceptual drawing showing the TPC drift volume. The axion would convert to a photon inside the TPC, interacting in the gas and producing electrons drifting towards the micromegas readout, allowing to measure time and spatial event topology. A proper shielding against external radiation should be placed in order to minimize the background level of the detector. }
\label{fi:amelieDetectorConcept}
\end{center}
\end{figure}

The new helioscope technique proposed in this section is based on an original idea developed in 1989, and described in~\cite{vanBibber}. In this work the authors present a stationary helioscope using a superconducting magnet, filled with a low-$Z$ buffer gas, like H$_2$ or He. The x-rays converted in the buffer gas would travel towards a thin window defining the boundary of the gas buffer, and would be then detected, for example, by a multiwire chamber. The novelty of the idea we present in this paper is the fact that the buffer gas and the x-ray detection volume would be the same entity. This could be made by means of a gaseous TPC immersed in an intense magnetic field in such a way that the detector itself performs axion-photon conversion and detects the photon absorbed. This would allow to increase the detection efficiency while using higher-Z gases. The TPC could be built using a micropattern readout; i.e., micromegas technology (detailed in section~\ref{sec:micromegas}). Micromegas detectors have shown good radiopurity and have proven the capability to operate in intense transverse magnetic fields ($\gtrsim$\,4\,T) for the CLAS12 experiment~\cite{clas121,clas122}. A TPC using a transverse magnetic field will produce a non-negligible deviation of the electron tracks drifting towards the micromegas read-out~\cite{procureur}, thus a natural choice is to use a magnetic field parallel to the drifting field in order to minimize the effect on the Lorentz angle  (see Figure~\ref{fi:amelieDetectorConcept}).

The fraction of axion-photon conversions that are directly absorbed by the gas was also detailed in~\cite{vanBibber}. The axion-photon absorption probability, $P^A_{a\gamma}$ is given by the following relation,

\begin{equation}\label{eq:absorption}
P^A_{a\gamma} = \frac{(g_{a\gamma}B/2)^2}{(q^2+\Gamma^2/4)^2} \Big\{ q^2 - 3\Gamma/4 + (q^2+\Gamma^2/4)(\Gamma L - e^{-\Gamma L}) + \left[ \Gamma^2 \mbox{cos}(qL) - 2q\Gamma \mbox{sin} (qL) \right] \times e^{-\Gamma L/2}  \Big\}
\end{equation}

\noindent which can be approximated for gases with high photo-absorption ($\Gamma L \gg 1$) by the following expression,

\begin{equation}
P^A_{a\gamma} = \frac{(g_{a\gamma}B/2)^2}{q^2+\Gamma^2/4} \Gamma L.
\end{equation}

\medskip
Therefore, for gases with high $\Gamma$ where the above approximation is valid, the axion absorption probability, $P^A_{a\gamma}$, will be higher than the transmitted photon component given by $P^T_{a\gamma}$, as it is observed by comparison with relation\,(\ref{eq:conversion}). The enhanced detection efficiency for gases with high photo-absorption coefficient implies this technique will be specially competitive for the higher axion mass region, $\sim$\,eV. 

\begin{figure}[ht!]
\begin{center}
\includegraphics[width=7.5cm]{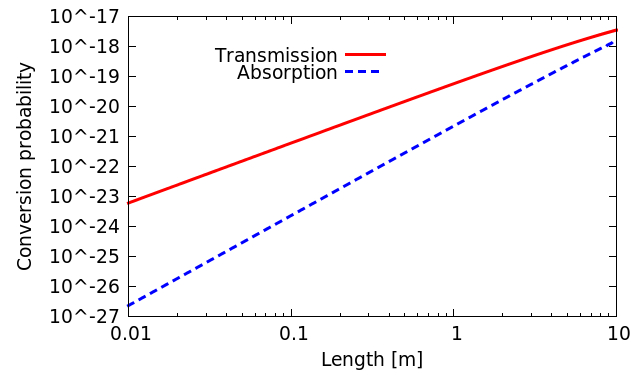}
\includegraphics[width=7.5cm]{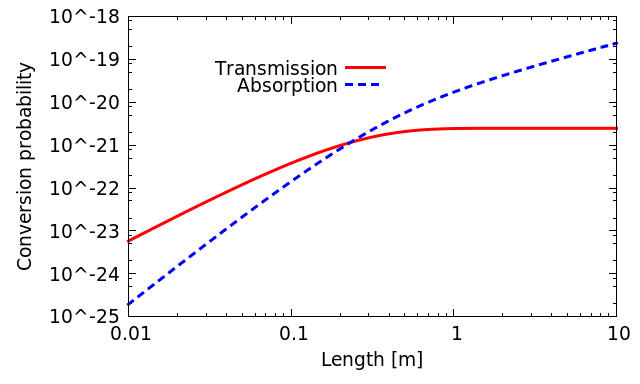}
\caption{ Axion-photon transmission probability, $P^T_{a\gamma}$, and absorption probability, $P^A_{a\gamma}$, as a function of the distance through a magnetic medium of 5\,T. On the left, helium at 10\,bar ($m_a\sim0.85$\,eV). On the right, xenon at 100\,mbar ($m_a\sim0.4$\,eV). }
\label{fig:photonTransmission}
\end{center}
\end{figure}

\medskip
For comparison, Figure~\ref{fig:photonTransmission} shows the absorbed and transmitted photon component produced by the axion flux traversing a magnetic field as a function of distance. The transmitted component saturates when the number of photons produced is comparable to the number of photons re-absorbed. The distance at which this occurs will strongly depend on the gas absorption coefficient and density.

We must notice that we will integrate the axion flux for a given detector section, $A$. Therefore, the quantity we must consider for optimizing detection sensitivity is $B^2V$, where $V$ is the volume of the detector.  Contrary to tracking helioscopes, our detection volume does not require to be a long pipe (geometry used to optimize the quantity $B^2VL$ derived from relation\,(\ref{eq:convVacuum}) after integration of the axion flux). Indeed, for searches based on detecting the absorbed photon component, a gaseous TPC volume with no privileged direction might be a better option. An obvious choice for implementation could be a cylindrical TPC such that $R=L/2$. In principle, the maximum amplitude of the axion transmission probability reachable is higher than in the case of the absorption probability\footnote{Due to the fact that in the absorption approach we must use higher $\Gamma$ gases}, however this is compensated by the possibility to use a larger magnet, increasing detection volume and exposure time. After all, this helioscope technique would be measuring the solar axion flux continously, with an efficiency that would modulate during the day due to the effective transversal component of the TPC magnetic field with respect to the incident axion flux.

%A particularity of this detector is that, obviously, it cannot operate in vacuum, so that lower axion cannot be accessed at full sensitivity (except that future development would allow operating the detector at very low pressures). This experiment will be leading in sensitivity, respect to the actual running searches, for axion masses above 10 meV.

\subsection{Annual and daily signal modulation}
\label{sec:modulation}

\begin{figure}[ht!]
\begin{center}
\includegraphics[width=7.5cm]{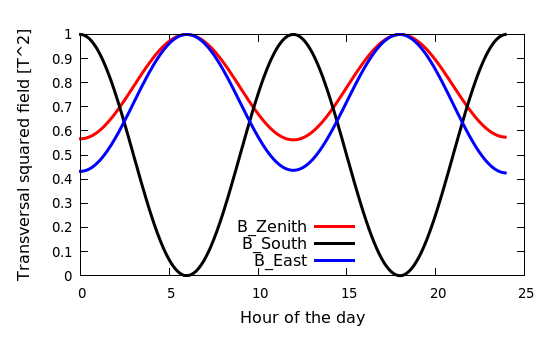}
\includegraphics[width=7.5cm]{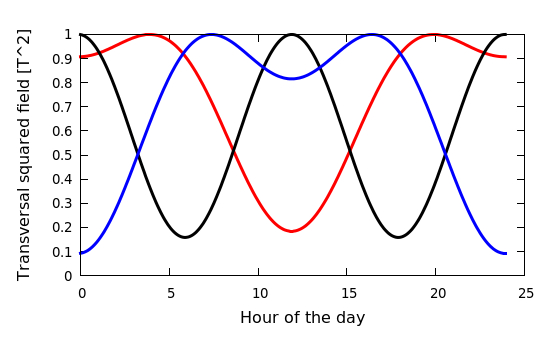}
\caption{ Daily evolution of the squared transversal field component for a field of $1\,T$, the 21st of March (left) and the 21st of June (right). The magnetic field is assumed to be homogeneous all throughout the TPC. Lines corresponding to three absolute magnetic field orientations are plotted, one being perpendicular to the ground ($B_{Zenith}$), the second in the North-South direction ($B_{South}$), and the last one in the East-West direction ($B_{East}$). Curves have been calculated for a latitude 48$^\circ$N and longitude 2$^\circ$E.  }
\label{fig:dailyMod}
\end{center}
\end{figure}

Using the proposed cylindrical geometry would allow to track the Sun at any time. The detector would be ideally installed stationary at an underground laboratory, being its axion-photon conversion probability correlated to the axion incident angle with respect to the magnetic field direction. If at a given moment of the day the detector is sitting such that the incident axion flux is fully transversal to the magnetic field the conversion would be maximized. If during the day, due to the rotation of the earth, the incident angle would vary between 0 and 45 degrees, it would produce a modulation signal varying between $B^2$ and $B^2/2$. Then, the axion flux would produce a time modulating pattern (see Figure~\ref{fig:dailyMod}) that would vary along the day, depending also on the period of the year. Therefore, the solar axion flux would produce an unmistakable signal in time. 

Another interesting feature of this helioscope type is the wide field of view, allowing to observe other celestial phenomena (i.e. supernovae) or unexpected axion sources. The daily modulation pattern will be unique for a specific position. Spatial resolution will strongly depend on statistics, something that could be improved if two or more detectors with different magnetic field orientations would be used. For example, if an excess would be observed during a short supernovae event, the detected event ratio between the different detectors would at least allow to validate its spatial origin. 

\medskip
Furthermore, the wide field of view would allow to measure the axion flux from the entire solar disk, adding the possibility to detect axions (or axion like particles) produced at the outer layers of the Sun or the limb. This capability is hardly reached by tracking helioscopes given their lower field of view imposed by the length of the magnet pipe and the angular acceptance of x-ray optics. One of the technical advantages of this helioscope is that not high accuracy alignment would be necessary since a systematic uncertainty on the TPC orientation would produce a small effect on the daily pattern and overall signal efficiency.

\begin{figure}[ht!]
\begin{center}
\includegraphics[width=12cm]{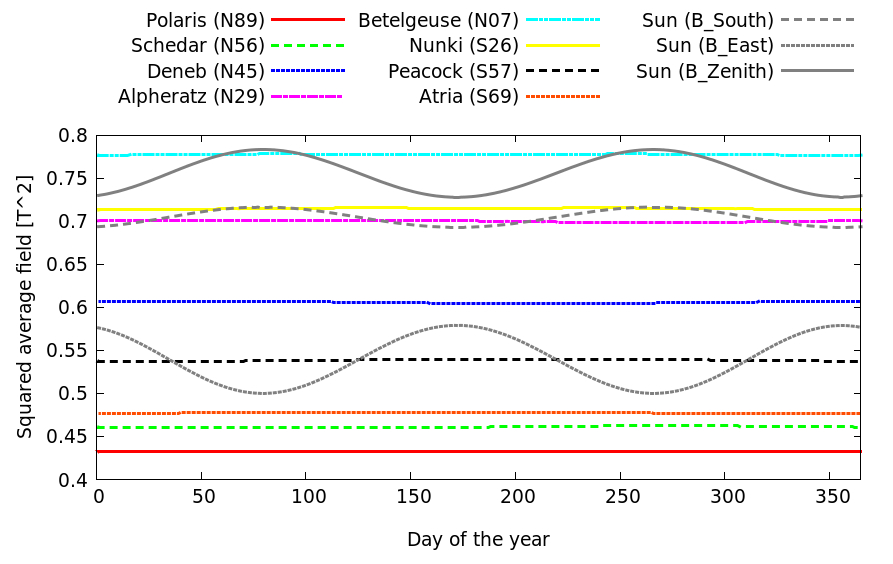}
\caption{ Annual evolution of the squared transversal field component (daily averaged) for 1\,T magnet and different positions in the sky by using navigation stars. The TPC magnetic field is defined pointing towards Zenith. The average transversal field component for the Sun is also shown for a magnetic field pointing towards South, East and Zenith. }
\label{fig:annualMod}
\end{center}
\end{figure}

\medskip
Figure~\ref{fig:annualMod} shows the squared transversal field component (daily averaged) along the year for different star locations. The relative orientation of the Earth respect to the Sun would produce a modulation of the detection efficiency as a function of the day of the year. For far away sources, the variation along the year is negligible and the daily detection efficiency would remain constant. As it is observed, the efficiency of detection will depend on the point of the sky a possible axion source would be coming from. From the three field orientations chosen for this figure (for the case of the Sun), the one pointing towards the Zenith is the most efficient for solar axions, reaching an average efficiency of about 75\%, equivalent to 18 hours of data taking per day keeping the magnet oriented towards the Sun. We must notice that this configuration, a TPC with a field perpendicular to the ground, is a natural choice in rare event physics searches with TPCs.

The daily and annual modulation patterns (Figure~\ref{fig:dailyMod} and Figure~\ref{fig:annualMod}) show just the modulation of the transversal field component. However, the overall detection efficiency will be affected by the fact that the axion trajectories inside the actual magnetic field geometry will depend on the incident angle and the impact parameter. Therefore, the conversion probability must be integrated to the different coherence lengths, $L$, for a given TPC geometry as a function of the incident angle. This leads to a geometrical factor that is more pronounced for lower-Z gases and lower densities (for trajectories where $\Gamma L \gg 1$ is not satisfied). We have calculated this geometrical factor in xenon at 10\,mbar, 50\,mbar and 100\,mbar for a cylindrical geometry ($R=L/2$) with its axis perpendicular to the ground obtaining an efficiency loss on conversion probability of 21.5\%, 11.9\% and 6.9\%, respectively. The effect of the transversal magnetic field component on the daily modulation is still dominant, and the effect of the geometrical factor for the higher pressure range becomes negligible.

%\begin{figure}[ht!]
%\begin{center}
%\includegraphics[width=9cm]{figures/convProbMod.jpg}
%\caption{ Daily evolution of the conversion probability as consequence of the geometrical factor mentioned in the text, related to the different coherence lengths covered by the TPC volume. We show the calculation for a cylindrical TPC  ($R=L/2$), filled with xenon at different pressures, and its axis perpendicular to the ground.  }
%\label{fig:convProbMod}
%\end{center}
%\end{figure}

%\begin{itemize}
%\item full view, no alignment required
%\item other objects under reach. No annual modulation for objects , only Sun. Field efficiency.
%\item several detectors for better coverage, detector network.
%\end{itemize}

%Additionally, other possible patterns could be observed due to exotic, or unexpected, axion sources which could be localized in space thanks to the relation between the incident axion flux and the movement of the earth.

\subsection{Resonance width and axion mass coverage}
%The final axion mass this helioscope will be sensitive to will depend on these two gas parameters, therefore the gas choice for a given axion mass, $m_a$, region must be studied in terms of axion coupling sensitivity, $g_{a\gamma}$. 

As in the case of tracking helioscopes using a buffer gas, the helioscope here proposed would be most sensitive to the axion mass matching the refractive photon mass of the medium, which depends on the gas pressure through relation\,(\ref{eq:mgamma}). The flexibility of operation of a TPC at low or high pressures, and different gas mixtures, will allow extending this search to a wide axion mass range. The lower pressure to be used will be ultimately determined by the electron track length which is finally detected after photon conversion in the TPC. If the track length of the electron starts to be comparable to the detector size it will imply a considerable reduction of effective detection volume. For instance, if the resulting electron track would be 5\,cm in a 1\,m$^3$ scale TPC, the active volume would be reduced to $\sim91$\% (in the worst scenario). Therefore, we should use a combination of gas mixture and pressure that keeps the track length below 5\,cm. The low-Z gases are the most affected ones, for He, Ne and Ar a track length of $\lesssim$\,5\,cm (for a 5\,keV photon) would imply to operate at a minimum pressure of 100\,mbar, 20\,mbar and 4\,mbar, respectively. While for Xe, a pressure as low as 4\,mbar already implies an electron track $\lesssim$\,5\,mm. The table below shows the axion mass range coverage, based on the refractive axion mass from relation\,(\ref{eq:mgamma}), for different gases and for the pressure range specified.

\begin{table}[h!]
\center
\begin{tabular}{c|c|c|c}
 Helium & Neon & Argon & Xenon \\
\hline
100\,mbar - 10\,bar & 20\,mbar - 10\,bar & 4\,mbar - 10\,bar & 4\,mbar - 10\,bar \\
 86\,meV - 0.86\,eV & 137\,meV - 1.94\,eV & 51\,meV - 2.57\,eV & 79\,meV - 4.05\,eV \\
\end{tabular}
\end{table}

The conversion probability for each helioscope technique is given by relations\,(\ref{eq:conversion}) and (\ref{eq:absorption}). Both relations describe a resonance centered at $m_\gamma$. These relations imply that the conversion resonance will be broadening for gases with higher $\Gamma$. Although the amplitude of the conversion probability is lower for higher $\Gamma$ values, the final sensitivity in a wide axion mass range will be roughly independent of the gas choice, since the integral of the resonance is not strongly dependent on $\Gamma$. Then, an advantage of using heavier elements as buffer gas would be to cover a wider axion mass region with fewer pressure steps, assuring additional stability during long data taking periods. Long data taking periods at each pressure setting are highly desirable, since a background measurement in absence of magnetic field would allow to probe an axion signal by direct comparison with data taken with a given magnetic field intensity.

\begin{figure}[htbp]
\begin{center}
\includegraphics[width=12cm]{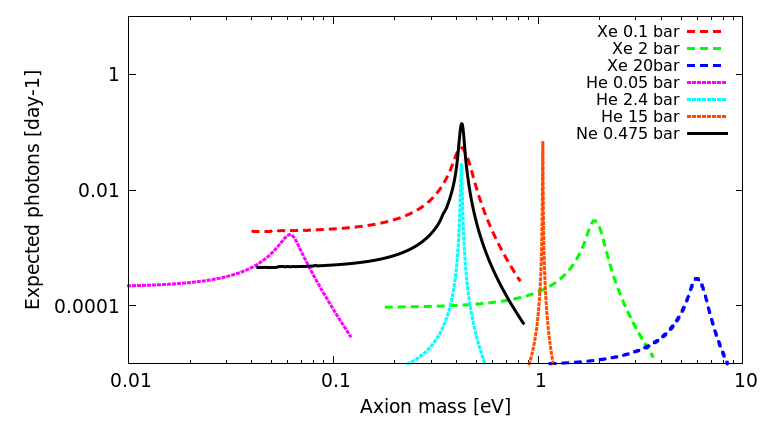}
\caption{Conversion probability resonance for different gas settings. Helium at 50\,mbar, 2.4\,bar and 15\,bar, neon at 475\,mbar, and xenon at 100\,mbar, 2\,bar and 20\,bar. The curves have been calculated for $B=1\,T$ and $V=1\,m^3$, by using relation\,(\ref{eq:absorption}).}
\label{fig:resonances}
\end{center}
\end{figure}

Figure~\ref{fig:resonances} shows the axion-photon conversion probability for different gases and pressures. It is observed that for the same $m_\gamma$, xenon shows broader axion mass coverage versus neon and helium. For instance, xenon could be used for a wide axion mass coverage, while neon could be used for increasing sensitivity at a particular axion mass. The amplitude of the resonance should in principle decrease with the axion mass, however for helium the effect of low photo-absorption is still dominant as pressure is increased.

\section{State of the art in low background large volume TPC's}
\label{sec:detector}

A fundamental issue for the success on the development of this technique resides on the background level achievable by a large volume TPC. In the tracking helioscope technique, the detector size requirements are just driven by the cold bore area, and in the case a focusing device is used, the signal to background ratio will be reduced to a small region of the detector. However, for the detection technique we propose the detector volume scales with the magnet size, and therefore so does its background level. Thus, a large volume TPC should achieve extraordinary good background levels to be competitive with tracking helioscopes.

There are several projects developing large volume TPCs for rare event searches. We can use the expectation on background level to be reached on those large volume prototypes, and use them as a reference to estimate the sensitivities reachable with the proposed helioscope technique. We will introduce in the following sections two R\&D projects under development for WIMP searches, describing their detection principle, advantages they present for our particular search (in terms of energy threshold and energy resolutiion), evolution and prospects on background levels. The first R\&D project we will describe is based on the development of a Spherical Proportional Counter (SPC), and the second is based on a large volume TPC using micromegas technology. The geometry of the electric field and magnetic field orientation will also play a role on the final choice for a particular TPC configuration. However, in this first approach we will focus on the background level achievable by this type of detectors. Further studies should determine the best TPC magnet setup for this kind of search.

\subsection{Spherical Proportional Counters}

\begin{figure}[htbp]
\begin{center}
\includegraphics[width=8cm]{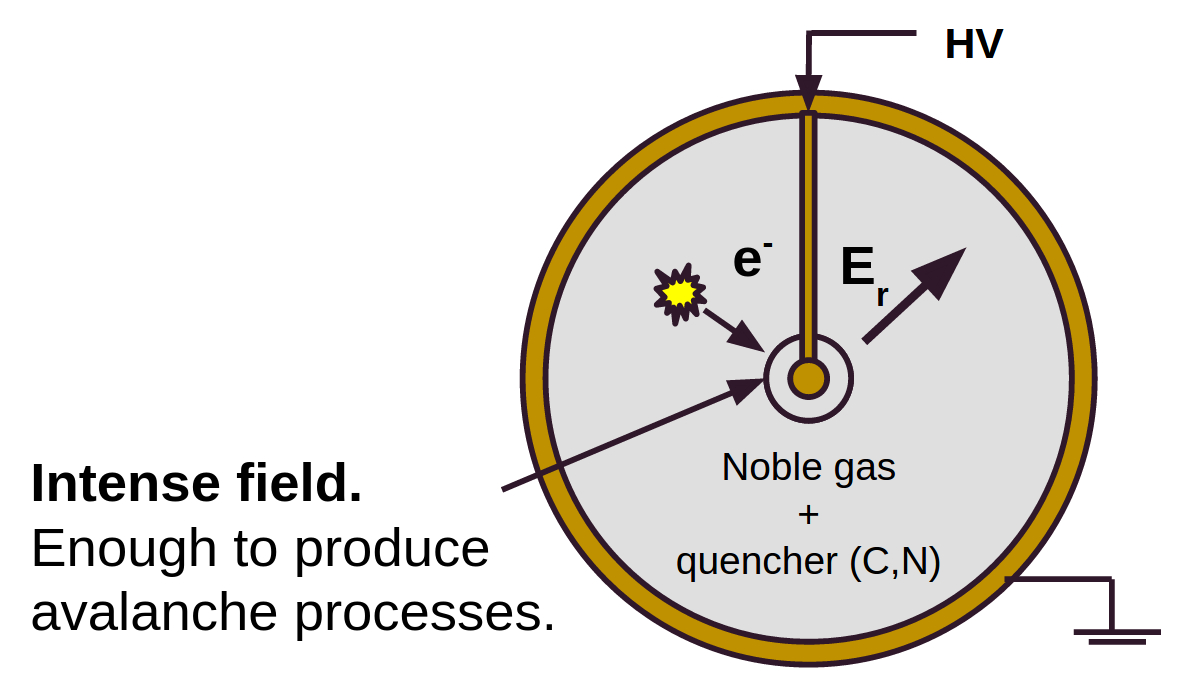}
\caption{A drawing describing the detection principle in a spherical TPC.}
\label{sphericalTPC}
\end{center}
\end{figure}

A Spherical Proportional Counter (SPC) consists of a spherical grounded cavity which is filled with gas~\cite{SPC1}. In the center of this cavity is placed a small spherical sensor (made of metallic or resistive material) where a high positive voltage is applied (see Figure~\ref{sphericalTPC}). The field produced inside the cavity allows drifting the charges produced by ionizing interactions in the gas. The field close to the sensor (typically 1\,cm diameter) is high enough to produce signal amplification through electron avalanche processes.

Recent work has shown the capability to obtain competitive background levels with a large volume TPC. A spherical TPC (SEDINE), with a volume of 110 dm$^3$, has been running at the Laboratoire Souterrain de Modane (LSM). SEDINE vessel has been built with radiopure copper. The internal radius is 30\,cm, and although different sensor types and sizes can be used, the detector is actually operating with 6.3\,mm diameter Silicon ball. Several interventions took place in the detector set-up to improve materials radiopurity and shielding design, allowing the reduction of the background level to 60 counts day$^{-1}$keV$^{-1}$bar$^{-1}$m$^{-3}$~\cite{SEDINE}. In addition, studies have shown that this background level is still limited by external gamma radiation and additional shielding upgrades and studies could help to reduce even further the actual values.

\medskip
SEDINE started operating at LSM in 2011. The detector is protected from external radiation by several layers of different materials, including radiopure copper, lead and polyethylene. Emphasis was placed on the radiopurity materials used and different cleaning processes have minimized the effect of contamination due to radon daughters and other contaminants sticking on the surrounding walls of the shielding and detector. The background levels obtained are competitive with the most sensitive experiments up to date for low mass WIMP searches~\cite{WIMPS1,WIMPS2,WIMPS3}. A near future program for SEDINE includes increasing target mass for gas pressures that could go up to 10 bar~\cite{NEWS}.  

\medskip
The main advantages of this detector reside on its simplicity, its large volume, energy resolution achievable (11\% FWHM at 6keV) and low electronic noise provided by the spherical topology, given by its low capacitance. The low energy threshold is only limited by the ionization energy of the gas~\cite{SPC2}. The discrimination capabilities provided by the time signal recorded allow to differentiate between local (x-rays) and extended charge depositions (electrons and muons).
%The dynamic range of the detector can be adjusted by using different values of the amplification field from few eV to several MeV, allowing us to scan interactions from low energy gammas to alphas or heavy ions.

\medskip
A single read-out channel allows acquiring the charge produced by the interacting particles which are drifted by the electric field towards the sensor, inducing a signal shape that depends on the charge density profile deposited in the gas and the distance to the sensor. The signal depends on the position through the drift time, given by the drift velocity of the electrons, and the charge diffusion (these quantities being also dependent on the electric field which varies along the radius of the sphere). The total drift time required by the resulting charges moving towards the sensor depends strongly on the gas mixture used and pressure (varying between few $\mu$s to several ms). The diffusion of electrons during their movement towards the sensor will depend on the distance to the sensor through the varying electric field. The strong dependence on the field as a function to the distance allows to exploit the charge diffusion to localize them and use this information to reduce surface event contamination.

\subsection{Micromegas-read TPC}
\label{sec:micromegas}
\begin{figure}[ht!]
\begin{center}
\includegraphics[width=8cm]{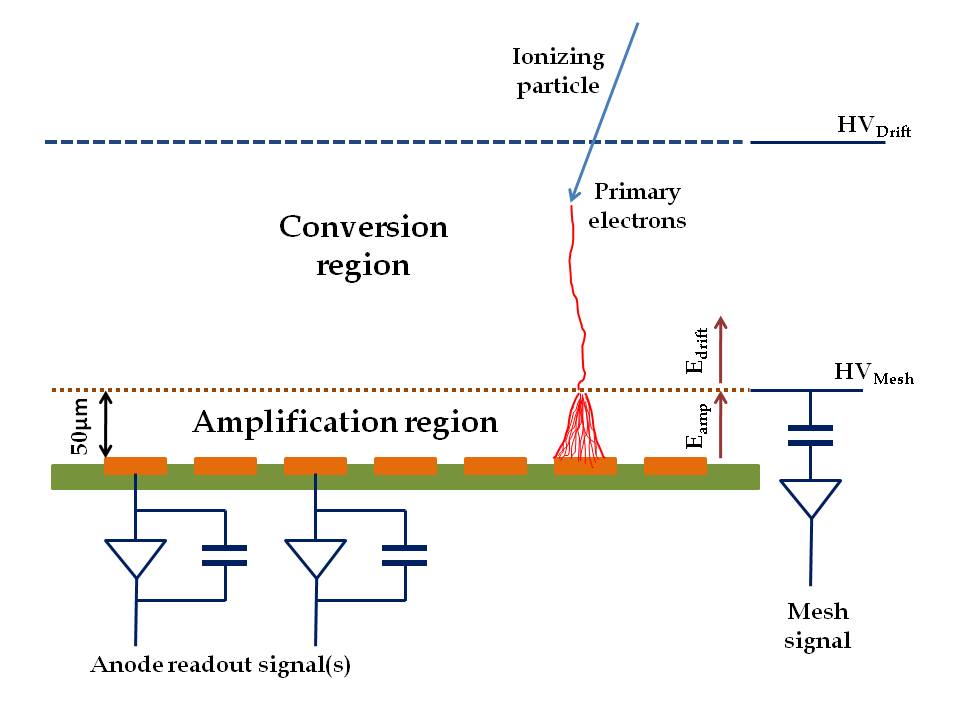}
\caption{Micromegas detector concept described in the text.}
\label{mmConcept}
\end{center}
\end{figure}

Micromegas (for MICRO MEsh GAseous Structure) is a parallel-plate detector invented by I.~Giomataris in 1995\,\cite{Giomataris:1995fq}. It consists of a thin metallic mesh and an anode plane, separated by insulating pillars. Both structures define a very little gap (between 20 and 300~$\mu$m), where primary electrons generated in the conversion volume (see Figure~\ref{mmConcept}) are amplified, applying moderate voltages at the cathode and the mesh. This type of readouts show good spatial, time and energy resolution, a stable gain along time and little mass budget.

\medskip
The two actual fabrication technologies (bulk \cite{Giomataris:2006yg} and microbulk \cite{Andriamonje:2010sa}) create all-in-one detectors, with improved homogeneity, and solving technical problems on mesh displacement observed on the first micromegas detectors produced\,\cite{Giomataris:1995fq}. In the first case, two photo-resistive layers with the right thickness are laminated with the anode readout circuit and the mesh, forming a single object. The supporting mesh pillars are formed using a mask with the pillar pattern and illuminating the piece with UV light, the rest of the material is removed by a chemical bath. Bulk detectors are quite robust and large areas can be built. In the second case, the raw material is a thin flexible polyimide foil with a thin (5\,$\mu$m) copper layer on each side. The mesh is etched out of one of the copper layers of the foil, and the amplification gap is created by removing part of the kapton by means of appropriate chemical baths and photolithographic techniques.  The amplification gap is more homogeneous and the mesh thickness is thinner than bulk detectors. For this reason, microbulk detectors have shown the best energy resolutions among micro-patterned readouts, as low as 11.2\% FWHM at 5.9\,keV in argon-isobutane, 10.5\% FWHM at 5.9\,keV in neon-isobutane \cite{Iguaz:2012fi} and 7.3\% FWHM at 22.1\,keV in xenon-TMA \cite{Cebrian:2013sc}.  They are also clean from the radioactivity point of view \cite{Cebrian:2011sc} but are less robust than bulk detectors and their maximum size is limited to $30 \times 30$ cm$^2$, due to the current equipment limitations.

\begin{figure}[ht!]
\begin{center}
\includegraphics[width=8cm]{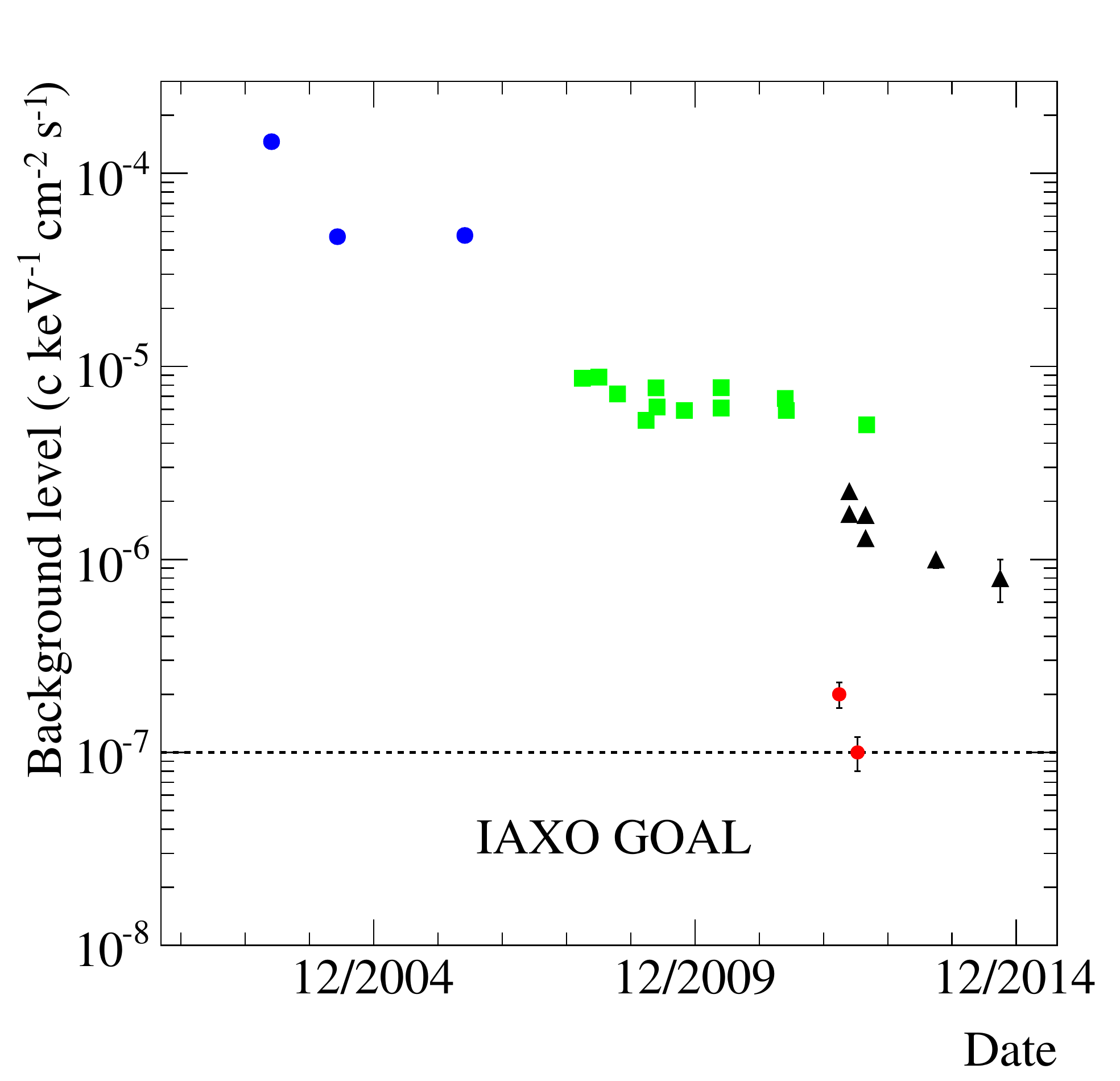}
\caption{Chronological improvements in background level with micromegas detectors at CAST for different detector set-ups (blue, green and black dots) and those achieved at the LSC using a micromegas detector and set-up similar to the one used in CAST (red dots)}
\label{bckTimeline}
\end{center}
\end{figure}

\medskip
Since micromegas detectors started operation at CAST~\cite{microCAST1,microCAST2,microCAST3,microCAST4}, several improvements on the detection line (i.e.; shielding upgrades, micromegas technology, etc) allowed lowering the natural background level measured, increasing the sensitivity of the experiment. Under coordination of the CAST group at the University of Zaragoza, background measurements took place at the Laboratorio Subterr\'aneo de Canfranc (LSC) with a radiopure micromegas detector\footnote{The same type of detector that was taking data in CAST.}~\cite{microLSC1,microLSC2,microLSC3}.   The aim of these measurements was to determine the background achievable with this technology. The set-up consisted of a faraday copper cage where the detector was placed, and a surrounding 20\,cm lead, 10\,cm polyethylene shielding. The results obtained showed that the detector in CAST was still not limited by the micromegas read-out materials, and that there was still room for optimization through shielding upgrade (see Figure~\ref{bckTimeline}).

The detector reached a background level of 10$^{-7}$ keV$^{-1}$ cm$^{-2}$ s$^{-1}$. We can express this background level in terms of the volume of the detector knowing that the detector drift was 3\,cm. Then, the renormalized background is about $3\cdot10^{3}$ keV$^{-1}$ m$^{-3}$ day$^{-1}$. However, for a large volume detector surface contamination effects would be minimized.

\begin{figure}[htb!]
\centering
\includegraphics[width=75mm]{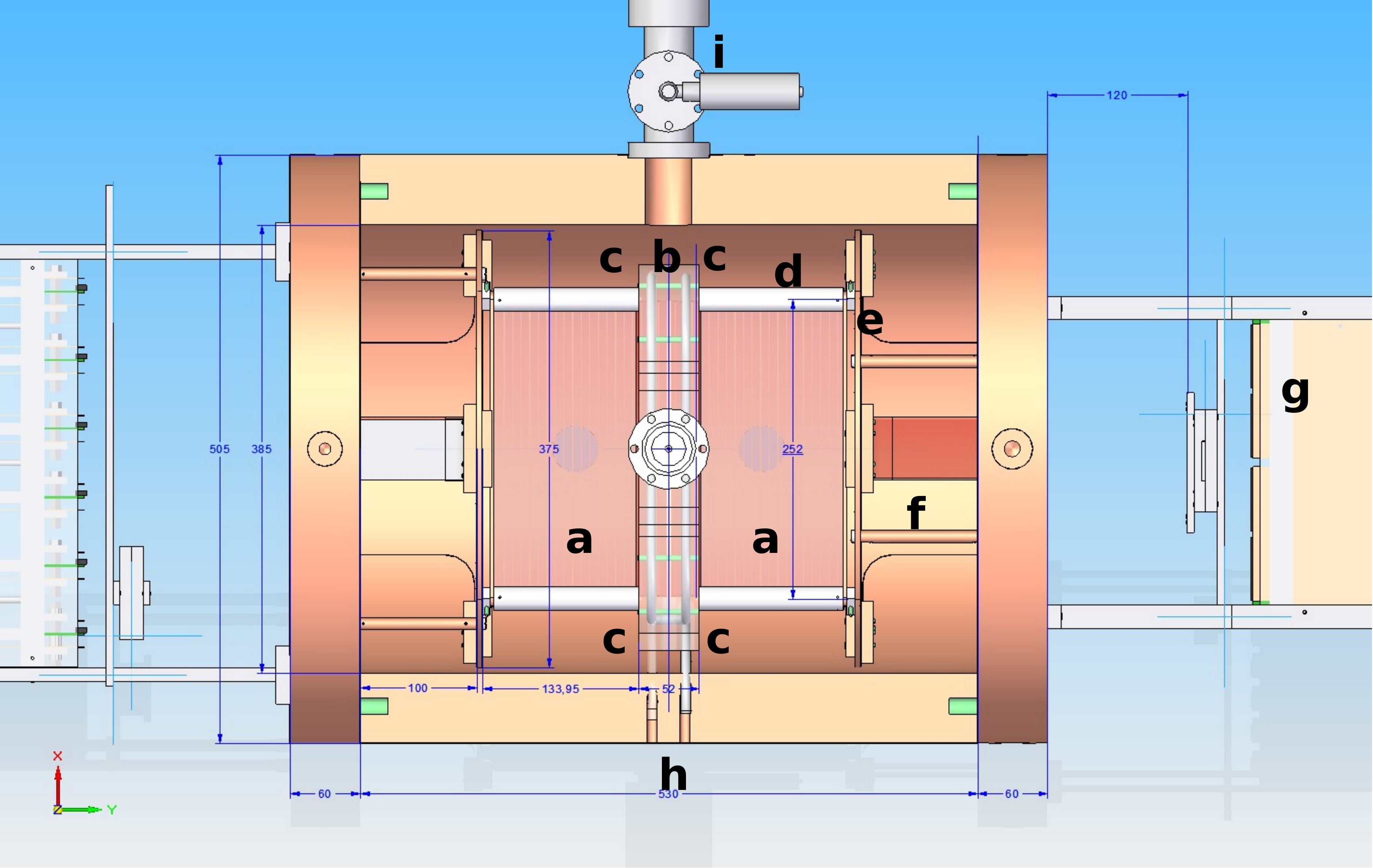}
\includegraphics[width=75mm]{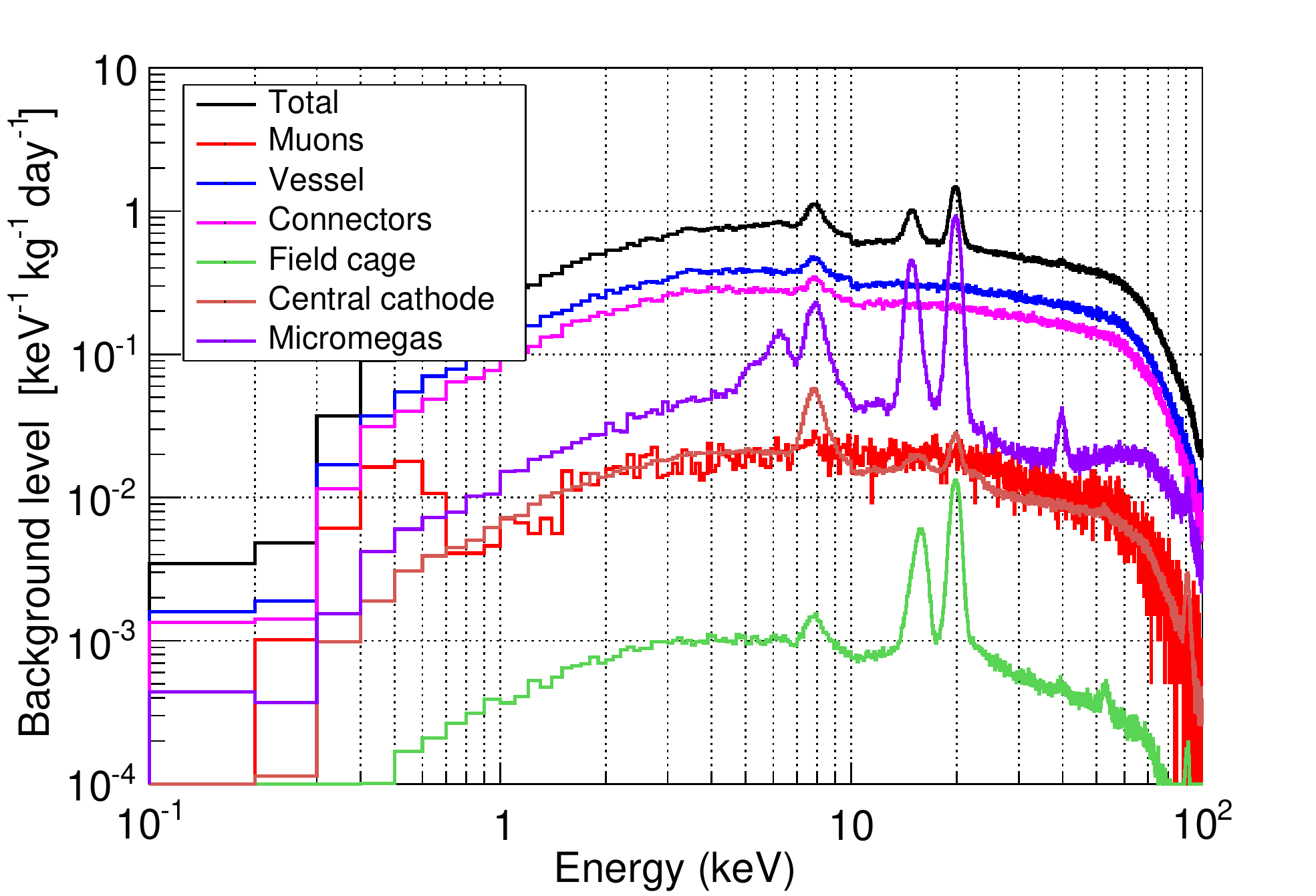}
\caption{Left: Design of the TREX-DM detector. The different parts are: active volumes (a), central cathode (b), calibration points (c), field cage (d), Micromegas detector and support base (e), flat cables (f), AFTER-based electronics (g), gas system (h) and pumping system (i).  Right: Background spectrum expected in TREX-DM experiment (black line) if operated in Ar+2\%iC$_4$H$_{10}$ at 10 bar in absence of any $^{39}$Ar isotope and installed at LSC. The contribution of the different simulated components is also plotted.}
\label{fig:Setup}
\end{figure}

\medskip
As a result of developments made in Micromegas technology \cite{Giomataris:1995fq, Giomataris:2006yg, Andriamonje:2010sa}, as well as in the selection of radiopure materials \cite{Cebrian:2011sc, Aznar::2013fa}, the University of Zaragoza is developing a low background Micromegas-based TPC for low-mass WIMP detection, called TREX-DM \cite{Iguaz:2015fi}.  Its main goal is the operation of an active detection mass $\sim$0.300 kg with an energy threshold below 0.4~keVee (as already observed in \cite{Aune:2014sa}) or lower. The actual setup (see Figure \ref{fig:Setup}, left) consists of a copper vessel divided into two active volumes, each of them equipped with a field cage and a bulk Micromegas $25 \times 25$ cm$^2$ bulk detectors. Signals are extracted from the vessel by flat cables and are read by an AFTER-chip based electronics. The actual setup is being commissioned in Ar+2\%iC$_4$H$_{10}$ up to 10 bar, and several improvements in detector's grounding are being made. In parallel to the commissioning, a large bulk Micromegas detector fully made of radiopure materials is being built.  This detector could be also read by AGET electronics~\cite{Anvar:2011sa}, which may further reduce the energy threshold to values near 100 eV. These improvements will be commissioned during 2015, so as the detector may be installed at the LSC during 2016 for a possible physics run.

\medskip

We estimated the background level achievable in TREX-DM experiment based on the Geant4 simulation of the radioactivity of the detector components, the signal response of a Micromegas-based TPC and a modified version of CAST analysis, used to discriminate low energy x-rays from muons. The expected background level (see Figure~\ref{fig:Setup}, right) in argon- and neon-based mixtures is 1-3 counts keV$^{-1}$ kg$^{-1}$ day$^{-1}$, i.e., 2.5 day$^{-1}$ m$^{-3}$ bar$^{-1}$ keV$^{-1}$.  These values must be confirmed by the simulation of the external gamma influence.

\section{Prospects for an AMELIE search}

\begin{figure}[htbp]
\begin{center}
\includegraphics[width=10cm]{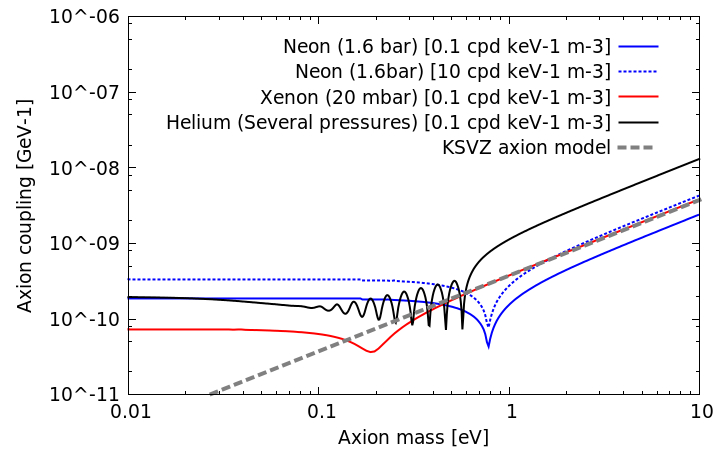}
\caption{ Sensitivity on the axion-photon coupling for the TPC conditions and data taking program described in the text, together with the theoretical coupling expected for KSVZ axion model. We show the sensitivity reachable with a single pressure step (1500 days exposure time) in neon (1.6\,bar) and xenon (20\,mbar). The sensitivity on neon has been calculated considering 2 different background levels 0.1 and 10 cpd\,keV$^{-1}$m$^{-3}$. We show also the combined limit of several pressure steps using helium as a buffer gas. }
\label{amelie_scenarios}
\end{center}
\end{figure}

We are now ready to show the final sensitivity of this type of helioscope by using the background level prospects for the large volume TPC projects presented in the previous section. We propose an Axion Modulation hELIoscope Experiment (AMELIE) exploiting the detection principle described in section~\ref{sec:modHeli}, using a large volume low background TPC immersed in an intense magnetic field. We have calculated the sensitivity reachable using different atomic elements as buffer gas (helium, neon and xenon), for a cylindrical TPC with an active volume given by $R=0.5$\,m and $R=L/2$, immersed in a magnetic field of 5\,T. We consider a background level of 0.1\,cpd\,keV$^{-1}$\,m$^{-3}$ for this calculation, based on the results and prospects for rare events searches with large volume TPCs. As it was mentioned, the final sensitivity of the experiment will depend on the background level achievable in our detector, thus we have considered also the conservative case of 10\,cpd\,keV$^{-1}$m$^{-3}$ in order to observe the effect on the final sensitivity of this type of helioscope. Figure~\ref{amelie_scenarios} shows different sensitivities reachable for different gas mixtures and conditions. Here we show the sensitivity for a 5\,years total exposure time (300 days per year) at two different pressures of neon and xenon, and for a pressure scanning using helium at different pressures varying between 10\,mbar and 5\,bar, and an exposure of 300 days per pressure step. An annual efficiency of 75\% on $B^2$ and a 20\% loss due to the geometrical factor (described in section~\ref{sec:modulation}) has been considered in the calculation.

%We consider a reasonable background level of 1\,cpd\,keV$^{-1}\,$m$^{-3}$ for this calculation. However, as it was mentioned, the final sensitivity will strongly depend on the background level achievable, thus we have also considered a conservative scenario and an optimistic scenario with a background levels of 10\,cpd\,keV$^{-1}\,$m$^{-3}$ and 0.1\,cpd\,keV$^{-1}\,$m$^{-3}$, in order to compare the effects of different background levels on the final sensitivity. Figure~\ref{amelie_scenarios} shows the prospects for a long data taking program using 11 different pressures steps, $P_n$, starting at 10\,mbar, $P_n=10$\,mbar$\times 2^n$. Each pressure step has been integrated over 300\,days period operating with magnetic field on, which could stand for a natural year data taking period considering some data is taken without magnetic field for detector background characterization in absence of magnetic field.

\medskip
As it was expected, xenon shows the best sensitivity in a wide axion mass range, due to its higher photo-absorption. Thus, for a first axion mass scan, using xenon as buffer gas seems the most reasonable choice. In case a signal would be observed, neon and helium could be used to reinforce the signal hypothesis and to gain in axion mass resolution, given the thinner axion mass resonance with lower-Z gases. Such sensitivity curves prove that a detector running in such conditions, using xenon as buffer gas, could allow to search for KSVZ axions above $\gtrsim$\,100\,meV.

%This data taking program would allow to proof KSVZ axions above $m_a\gtrsim 50$\,meV if neon would be used, or to reach enhanced sensitivity using xenon for the higher axion mass region $m_a\gtrsim$\,eV. Helium as buffer gas would allow to be sensitive to KSVZ axions on the axion mass range $25\,-50$\,meV. However, a longer exposure and more numerous pressure steps would be required. In spite of with helium is harder to cover a broad axion mass range using this detection technique, if the axion mass would be known and in this range, it would still be reachable by AMELIE. Thus, it could verify a possible future discovery in this axion mass region (i.e. if IAXO would find a positive signal in this region).

\begin{figure}[ht!]
\begin{center}
\includegraphics[width=10cm]{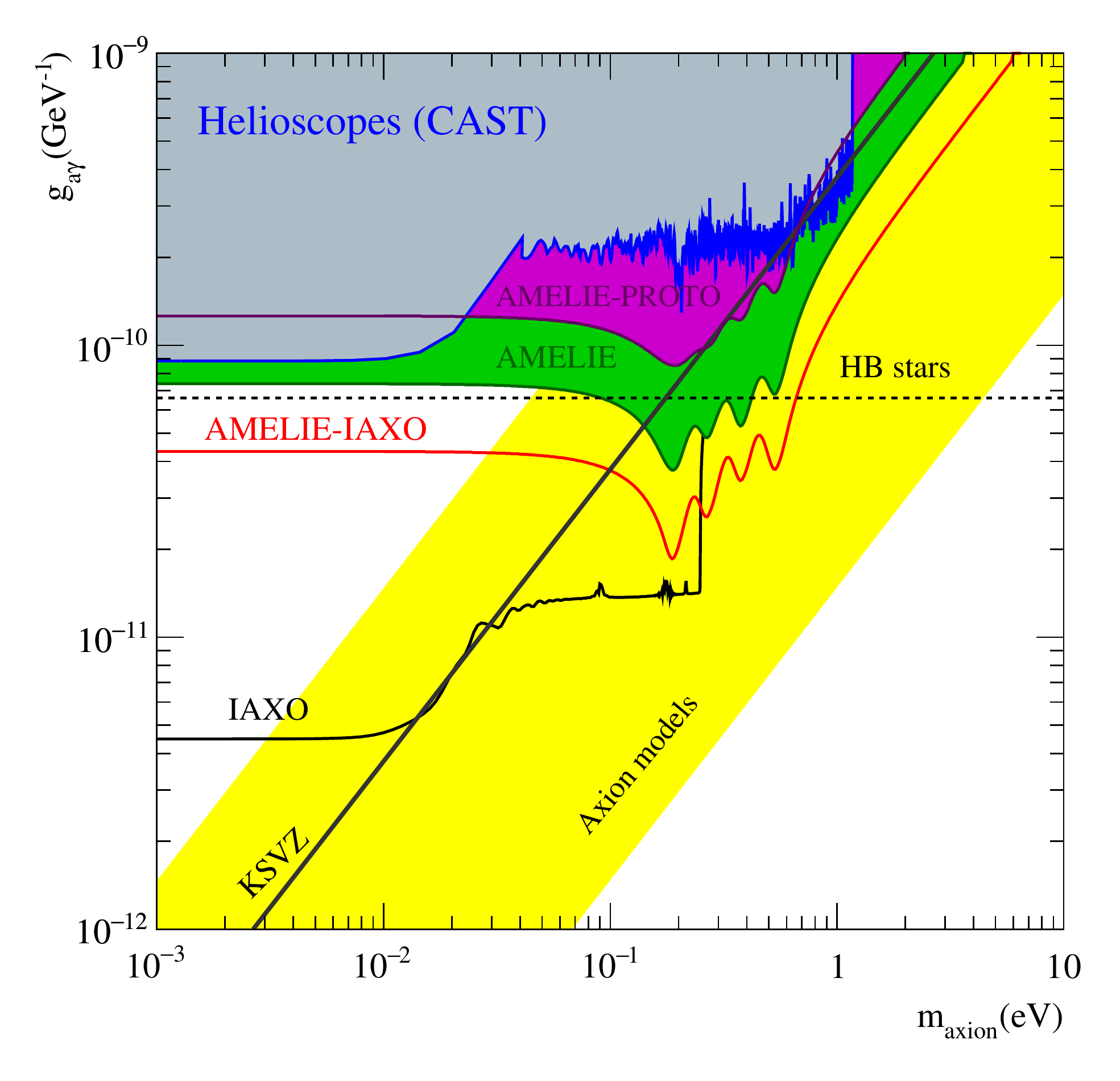}
\caption{ Axion-photon coupling as a function of the axion mass excluded by tracking helioscopes (CAST). Together with future IAXO sensitivity prospects. We plot the prospects for a small prototype (AMELIE-PROTO), a 1\,m$^3$ scale detector (AMELIE), and a hypothetical search using the characteristics of IAXO magnet (AMELIE-IAXO), for the pressure conditions and exposure described in the text. The 75\% efficiency due to the field modulation, and a 20\% efficiency loss due to the geometrical factor have been included. The yellow band represents the favored axion theoretical region. }
\label{amelie_IAXO}
\end{center}
\end{figure}

%AMELIE-PROTO : 1cpd/m3/keV  - 150 days per step - 16 Pressure steps (10mbar to 4bar) - 5T magnet
\medskip
The sensitivities shown would be reachable with a medium size TPC. We have calculated also the implementation of this detection technique in a magnet as IAXO. The IAXO effective magnet volume is $\sim$45\,m$^3$, increasing considerably the sensitivity of this detection technique. Figure~\ref{amelie_IAXO} shows the prospects for small size prototype (AMELIE-PROTO $\sim$\,21\,dm$^3$), a medium size TPC (AMELIE - same size as the prospects shown in Figure~\ref{amelie_scenarios}), and a hypothetical AMELIE-IAXO search. We define \emph{four} pressure settings at 20\,mbar, 40\,mbar, 80\,mbar and 160\,mbar using xenon as buffer gas. The total exposure time of the first pressure step is 5\,years, the second is 2.5\,years and the \emph{two} last ones are 1.25\,years (1\,year being 300 effective days).

We observe that a first search using this technique with a small scale prototype (AMELIE-PROTO) could already scan a region of the axion parameter space not explored by previous experimental axion searches. And, a medium size scale (AMELIE) could allow to probe KSVZ axions for masses above $\gtrsim$ 100\,meV, and allowing to go beyond the astrophysical limits derived from HB stars. As it is observed this technique would allow to extend the IAXO future sensitivity towards the higher axion mass region. If no axion signal would be found with IAXO, a second phase (i.e. placing IAXO underground) using this technique could enhance the IAXO sensitivity towards the higher mass axion region. 

\section{Conclusions}

We have presented a novel detection concept for the search of solar axions that have not been exploited so far in axion helioscopes. We have shown that the progress on low background large volume TPCs could allow to reach axion-photon coupling sensitivities improving running helioscope searches, and allowing to explore a new region in the theorical favored axion parameter space, not accessible before in previous axion searches.

\medskip
We summarize some of the advantages this axion helioscope technique would entail

\begin{itemize}
\item The helioscope would be placed stationary. This, together with the possibility to use a broad axion mass coverage (using higher-Z gases) would allow to increase stability. The data taking process would be simplified for long periods of time, since a broad axion mass range could be covered with a single pressure setting.
\item There is no high-accuracy alignment required and a full angular scan would be possible, allowing to measure the full solar disk, making possible to detect possible axion phenomenology related to the intense magnetic fields on the outer layers of the Sun.
%\item This technique would be an independent measurement method that could allow to reinforce a possible signal from any other experiment if the axion would be found in the axion mass range we would be sensitive to. 
\item If a signal would be found, the simplicity of design would allow to scale the experiment to a world wide network allowing to better characterize solar axions coming from the Sun and other possible axion sources, given the full angular view capabilities of this helioscope technique.
\end{itemize}

The development of this technique is motivated by the prospects on large volume TPCs for rare events searches presented in this work. To achieve the sensitivities we have shown it will be required to prove that we could achieve such background levels in a TPC when immersed in a magnetic field. Therefore, further studies on discrimination capabilities using a magnetic field should be performed. In principle, it is not obvious if the presence of a magnetic field would increase or reduce discrimination potential. We should also emphasize that progress on rare event searches during the last years has been pushing the background levels achieved even further, by exploiting the discrimination potential of the event topology, by improving the radiopurity of materials used in the detector, and by optimizing the structure of the shielding used to diminish the effect of external radiation.

\acknowledgments
%Y. Semertzidis
We acknowledge support from the European Commission under the European Research Council T-REX Starting Grant ref. ERC-2009-StG-240054 of the IDEAS program of the 7th EU Framework Program.
%This is the most common positions for acknowledgments. A macro is
%available to maintain the same layout and spelling of the heading.

%\paragraph{Note added.} This is also a good position for notes added
%after the paper has been written.


\end{document}